\documentclass[12pt,a4paper,final]{iopart}

\usepackage{iopams}
\usepackage{graphicx}
\usepackage[breaklinks=true,colorlinks=true,linkcolor=blue,urlcolor=blue,citecolor=blue]{hyperref}
\usepackage{appendix}

\begin{document}

\title[Many-Body Quantum Synchronisation]{Dynamics of Many-Body Quantum Synchronisation}

\author{C. Davis-Tilley, C. K. Teoh and A. D. Armour}
\address{Centre for the Mathematics and Theoretical Physics of Quantum Non-Equilibrium Systems and School of Physics and Astronomy, University of Nottingham, Nottingham NG7 2RD, United Kingdom}
\ead{andrew.armour@nottingham.ac.uk}

\begin{abstract}
We analyse the properties of the synchronisation transition in a many-body system consisting of quantum van der Pol oscillators with all-to-all coupling using a self-consistent mean-field method. We find that the synchronised state, which the system can access for oscillator couplings above a critical value, is characterised not just by a lower phase uncertainty than the corresponding unsynchronised state, but also a higher number uncertainty. Just below the critical coupling the system can evolve to the unsynchronised steady state via a long-lived transient synchronised state. We investigate the way in which this transient state eventually decays and show that the critical scaling of its lifetime is consistent with a simple classical model.
\end{abstract}

\section{Introduction}

The spontaneous synchronisation of limit-cycle oscillators\,\cite{acebron2005,Pikovsky2003} is a fascinating example of a phase transition which occurs far from equilibrium. Limit-cycle oscillators have a non-zero average amplitude, but no preferred phase, and are extremely common in both the physical and biological sciences. Synchronisation has been studied in a wide range of classical systems\,\cite{Pikovsky2003}, as well as in systems such as lasers where semiclassical descriptions prove accurate\,\cite{cresser,fabiny}. Over the last few years experiments have begun to investigate synchronisation in smaller-scale oscillator systems, including micron-sized mechanical oscillators\,\cite{Zhang2012,Bagheri2013,Matheny2014,Zhang2015} and lasers operating in the few-photon regime\,\cite{schlottmann}. Recently, theorists have also started to explore synchronisation in oscillators where a fully quantum mechanical description becomes essential\,\cite{zhirov,fazio13,ludwig,lee,zambrini,walter14,lee2,mendoza,holland,Zhu2015,fazio,Hush2015,weiss,brandes,li,davis,lorch16,lorch17,rz,sonar}.

Several different ways of quantifying the synchronisation of quantum oscillators have been proposed\,\cite{fazio13,lee,fazio,Hush2015,li} and the connection between synchronisation and entanglement\,\cite{fazio13,zambrini,lee2,fazio,li,davis,roulet1} has been examined in a variety of different systems.
Comparisons of synchronisation in quantum and semiclassical oscillator models have revealed significant quantitative and qualitative differences in behaviour\,\cite{lee,davis,lorch16,lorch17}. Detailed proposals have also been made for experiments which could probe synchronisation in the quantum regime using trapped ions\,\cite{lee,Hush2015}, optomechanical systems\,\cite{ludwig} or superconducting circuits\,\cite{nigg}.

Synchronisation in quantum models of many coupled limit-cycle oscillators\,\cite{ludwig,lee,brandes,kanamoto} has so far received rather less attention than few-oscillator systems. However, quantum many-oscillator models form a novel class of many-body system and the synchronisation transition they undergo makes an interesting comparison not just with classical or semiclassical oscillator systems, but also with the rich variety of non-equilibrium transitions which have been studied extensively in other types of many-body quantum systems\,\cite{noh,leecross,Marcuzzi2014,Schiro2016}, including e.g., the driven dissipative Bose-Hubbard model\,\cite{tomadin,diehl,ciuti1}.

A particularly simple model system for studying many-body synchronisation consisting of coupled quantum van der Pol oscillators was introduced by Lee and Sadeghpour\,\cite{lee}. In their work Lee and Sadeghpour compared the predictions of quantum and semiclassical versions of the model, and found that the transition which the oscillators undergo between unsynchronised and synchronised states consistently occurs at a lower value of the inter-oscillator coupling strength in the quantum case. Here we examine the properties of the synchronised and unsynchronised states of this system together with its critical dynamics. We find that the phase ordering which occurs when the oscillators synchronise is accompanied by other changes in the state of the system: a decrease in the average occupation number and an increase in number uncertainty. For coupling strengths just below the transition, the system displays critical slowing down with a long-lived transient synchronised state emerging and then eventually decaying. We look at exactly how the long-lived transient state decays and show that the critical scaling fits a simple classical model.

The rest of this paper is organised as follows. We begin by introducing the many van der Pol-oscillator model in section \ref{sec:vdp} and then go on to describe the synchronisation transition it undergoes in section \ref{sec:transition}. We compare the quantum properties of the synchronised and unsychronised states of the system in section \ref{sec:flucs}. We explore the dynamics close to the transition in section \ref{sec:unlocking} and then analyse the critical scaling in section \ref{sec:criticalscaling}. Finally, we present our conclusions in section \ref{sec:conclusion}.

\section{Model Oscillator System}
\label{sec:vdp}

The van der Pol (vdP) oscillator is a simple limit-cycle oscillator and is a popular choice of model to study synchronisation\,\cite{acebron2005,Pikovsky2003,Kuznetsov2009}. In the quantum regime the vdP model is described by a harmonic oscillator which gains individual photons at a rate $\kappa_1$ (linear anti-damping) whilst also losing two photons at a time with rate $\kappa_2$ (non-linear damping)\,\cite{lee}. The master equation in the interaction picture takes the form,
\begin{eqnarray}
\dot{\rho}&=&\mathcal{L}_d\rho\\
&=&\kappa_1(2a^{\dagger}\rho a-aa^{\dagger}\rho-\rho aa^{\dagger})+\kappa_2(2aa\rho a^{\dagger}a^{\dagger}-a^{\dagger}a^{\dagger}aa\rho-\rho a^{\dagger}a^{\dagger}aa) \label{eq:vdp}
\end{eqnarray}
where $a$ is the oscillator lowering operator.

The ratio $R=\kappa_2/\kappa_1$ controls the size of the limit cycle in the system\,\cite{lee,lorch16,exact}. For extremely small values of $R$ the average photon number becomes very large, growing as $1/R$, and semiclassical methods\,\cite{lee} provide an accurate description. However, for larger $R$ values  the behaviour of quantum and semiclassical versions of the model become markedly different\,\cite{lee} and eventually, in the limit $R\rightarrow \infty$ the steady state of the system becomes entirely confined to just the two lowest number states.

We are interested in the behaviour of an ensemble of $N$ identical vdP oscillators with all-to-all coherent couplings of strength $\varepsilon$, for which the Hamiltonian is
\begin{equation}
H_{int}=\hbar\frac{\varepsilon}{N}\sum_{i<j}\left(a_ia_j^{\dagger}+a_i^{\dagger}a_j\right).
\end{equation}
As is typically the case with quantum many-body problems, the extremely large state space involved precludes an exact numerical treatment and so we follow Lee and Sadeghpour\,\cite{lee} in assuming $N$ is large and adopting an approximate self-consistent mean-field approach\,\cite{tomadin,diehl,ciuti1,ludwig,lee}: we replace $a_ia^{\dagger}_j\rightarrow\langle a_i\rangle a_j^{\dagger}+a_i\langle a^{\dagger}_j\rangle$. This leads to an effective single-oscillator master equation for the system
\begin{equation}
\dot{\rho}=-\frac{i}{\hbar}[H_{mf},\rho]+\mathcal{L}_d\rho \label{eq:mfme}
\end{equation}
with the Hamiltonian
\begin{equation}
H_{mf}=\hbar\varepsilon\left(\langle a\rangle a^{\dagger}+\langle a^{\dagger}\rangle a\right).
\end{equation}
The master equation is solved self-consistently via numerical integration, starting from a chosen state, with the value of $\langle a\rangle$ updated at each time step\,\cite{ludwig}. In each case we carried out integrations of the master equation using a Runge-Kutta algorithm and generally chose a coherent state as an initial condition. We worked in the number state basis, using a cut-off, $nmax$, chosen to be large enough not to influence the results.

The self-consistent mean-field approach is very commonly used in studies of non-equilibrium quantum many-body systems\,\cite{noh,leecross,Marcuzzi2014,Schiro2016,tomadin,diehl,ciuti1}, often as an approximate description for a lattice of optical or microwave cavities in which individual cavities are coupled to a small number of their nearest neighbours. In such situations mean-field calculations provide a useful starting point though they are not expected to describe the behaviour faithfully in low-dimensional systems\,\cite{noh,jaksch,Chaikin1995}. In this case we have in mind a large number of individual vdP oscillators with all-to-all coupling and so expect that the mean-field approach will work increasingly well as that number is increased. Systems of many non-linear quantum oscillators with all-to-all couplings are of interest in other contexts and detailed proposals have been made for schemes which could realise such systems\,\cite{alltoall}.

Note that even after we have assumed a large number of coupled vdP oscillators, the value of the damping rate ratio $R$ can still be varied. This parameter controls the size of the {\emph{individual}} oscillators with $R\rightarrow 0$ in a sense setting both the `thermodynamic' and semiclassical limit for an uncoupled oscillator\,\cite{Casteels2017}.

\section{Synchronisation Transition}
\label{sec:transition}

 The long-time state of the many-body vdP system given by equation \ref{eq:mfme} displays either synchronised or unsynchronised behaviour depending on the coupling strength $\varepsilon$, the ratio of rates of the 2- and 1-photon processes, $R$, and the initial state of the system. The synchronised state is characterised by the emergence of a clear phase preference, signalled by a non-zero value of\,\cite{ludwig,lee} $\langle a\rangle$, which oscillates periodically in time with a magnitude that settles down to a constant value. In contrast, the  unsynchronised state has no preferred phase so that $\langle a\rangle=0$, leading to a time-independent state which matches the steady state of the corresponding uncoupled vdP oscillator. The value of $|\langle a\rangle|$ therefore provides a natural order parameter for the system.

  Initial states always exist which allow the system to reach an unsynchronised state, but the synchronised state can only be accessed for couplings beyond a certain critical value $\varepsilon_c$ which depends on $R$. The behaviour of $\varepsilon_c$ as a function of $R$ was mapped out by Lee and Sagedhpour\,\cite{lee}. The value of $\varepsilon_c$ vanishes in the limit $R\rightarrow 0$, grows rapidly with increasing $R$, before apparently diverging at a finite value of $R$. The behaviour of the vdP oscillator becomes more strongly quantum mechanical as $R$ is increased and this is reflected in the value of the critical coupling\,\cite{lee} which always takes a lower value in the quantum model compared to its semiclassical counterpart with the difference between the two growing rapidly with $R$.

\begin{figure}
	\centering
		\includegraphics[width=\textwidth]{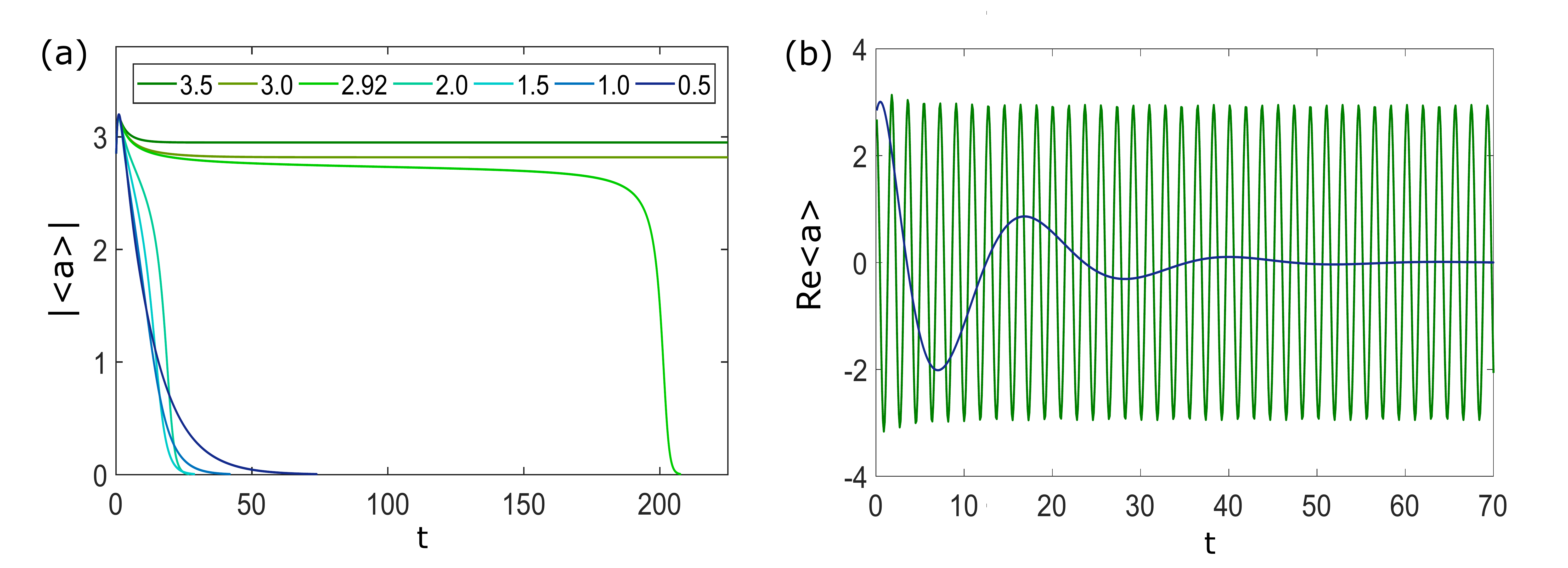}
\caption{Examples of the evolution of (a) $|\langle a \rangle|$ and (b) ${\rm Re}\langle a\rangle$ with time for different coupling values ranging from $\varepsilon=0.5$ (dark blue) up to $\varepsilon=3.5$ (dark green);  in each case $R=0.04$ and the numerical integrations used an initial coherent state with eigenvalue $\alpha=2.75$. The long-time value of $|\langle a\rangle|$ is non-zero for $\varepsilon>\varepsilon_c\simeq2.932$, signifying a synchronised state. We adopt units of time such that $\kappa_1=1$ throughout.}
\label{fig:tint}
\end{figure}

  Figure \ref{fig:tint} shows examples of the dynamics of $|\langle a\rangle|$ and $\langle a\rangle$ as a function of $\varepsilon$ with everything else kept fixed.  For $\varepsilon>\varepsilon_c$, $\langle a\rangle$ rapidly reaches a periodically oscillating state whose amplitude and period depend on $\varepsilon$. For the smallest values of $\varepsilon$ the value of $|\langle a\rangle|$ decays exponentially,
  but for values of $\varepsilon$ just below $\varepsilon_c$ the character of the decay is very different and it instead occurs in two distinct stages: a slow part followed by a very rapid part. The development of a very slow relaxation time in the dynamics is a precursor to the emergence of the synchronised state at $\varepsilon=\varepsilon_c$ and we investigate its properties in detail in sections \ref{sec:unlocking} and \ref{sec:criticalscaling} below. However, this is not the only interesting feature in the dynamics: for fairly weak couplings there is a regime in which the decay of the order parameter actually speeds up with increasing couplings [see figure \ref{fig:tint}(a)].

\begin{figure}
	\centering
		\includegraphics[width=\textwidth]{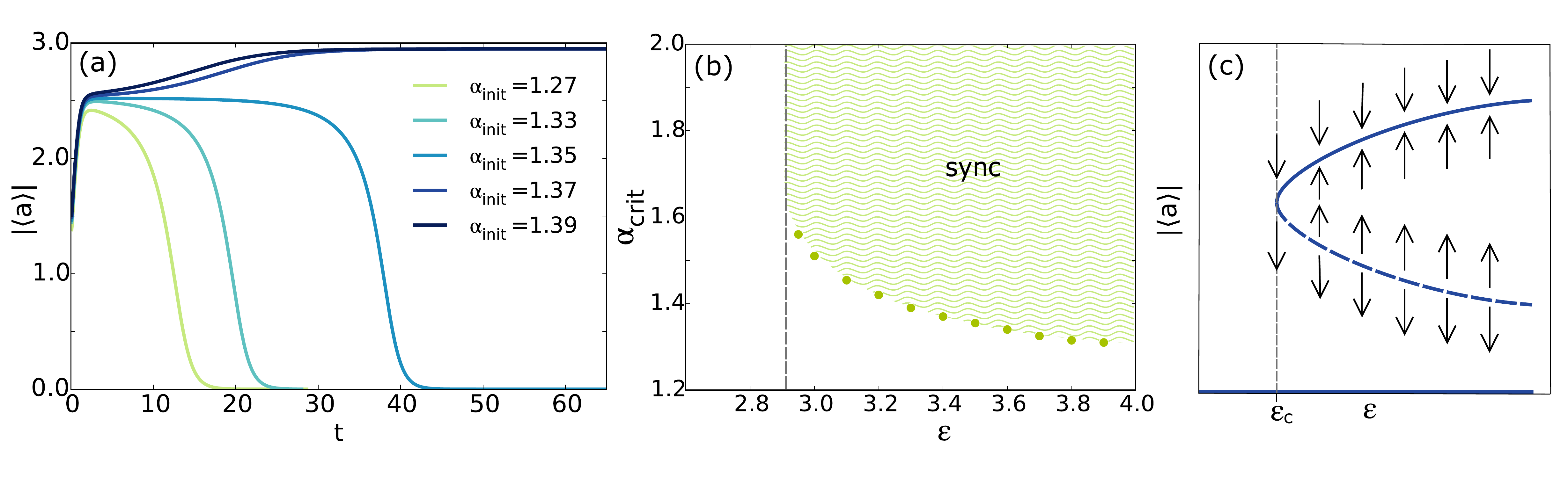}
\caption{(a) Evolution of $|\langle a \rangle|$ at $\varepsilon=3.5$ for initial coherent states with a range of eigenvalues, $\alpha_{\rm{init}}$, and  $R= 0.04$.  (b) Critical values of $\alpha_{\rm{init}}$ (green points) which lie on the separatrix between synchronised and unsynchronised states as a function of $\varepsilon$. The region where synchronised states arise is shown schematically (green shading) together with a (dashed) vertical line indicating the value of $\varepsilon_c$. (c) Sketch summarising the dependence of the behaviour on initial conditions. For $\varepsilon>\varepsilon_c$ a synchronised state (solid curve) emerges together with a separatrix (dashed curve) marking out the basins of attraction of the synchronised and unsychronised states in terms of initial values of $|\langle a\rangle|$. Arrows indicate the evolution in time for initial states in different regions. }
\label{fig:vdpinit}
\end{figure}

   When the coupling exceeds the critical value, $\varepsilon>\varepsilon_c$, the behaviour of the system in the limit of long times is determined by the initial state of the system. Using an initial coherent state with an eigenvalue, $\alpha_{\rm{init}}$, which is varied, we find a transition from unsynchronised to synchronised final states at a critical value, $\alpha_{\rm{crit}}$, as shown in figure \ref{fig:vdpinit}(a). The critical value of $\alpha_{\rm{init}}$ decreases with increasing $\varepsilon$, mapping out a separatrix between initial conditions that lead to synchronised and unsynchronised states as shown in figures \ref{fig:vdpinit}(b) and \ref{fig:vdpinit}(c). We note that although a dependence of the long-time density operator on initial conditions is not usually expected for open quantum systems, such behaviour does emerge when they are treated approximately, using e.g. self-consistent mean-field methods, as we do here\,\cite{Marcuzzi2014,ciuti1,jaksch}.

\begin{figure}
	\centering
		\includegraphics[width=0.7\textwidth]{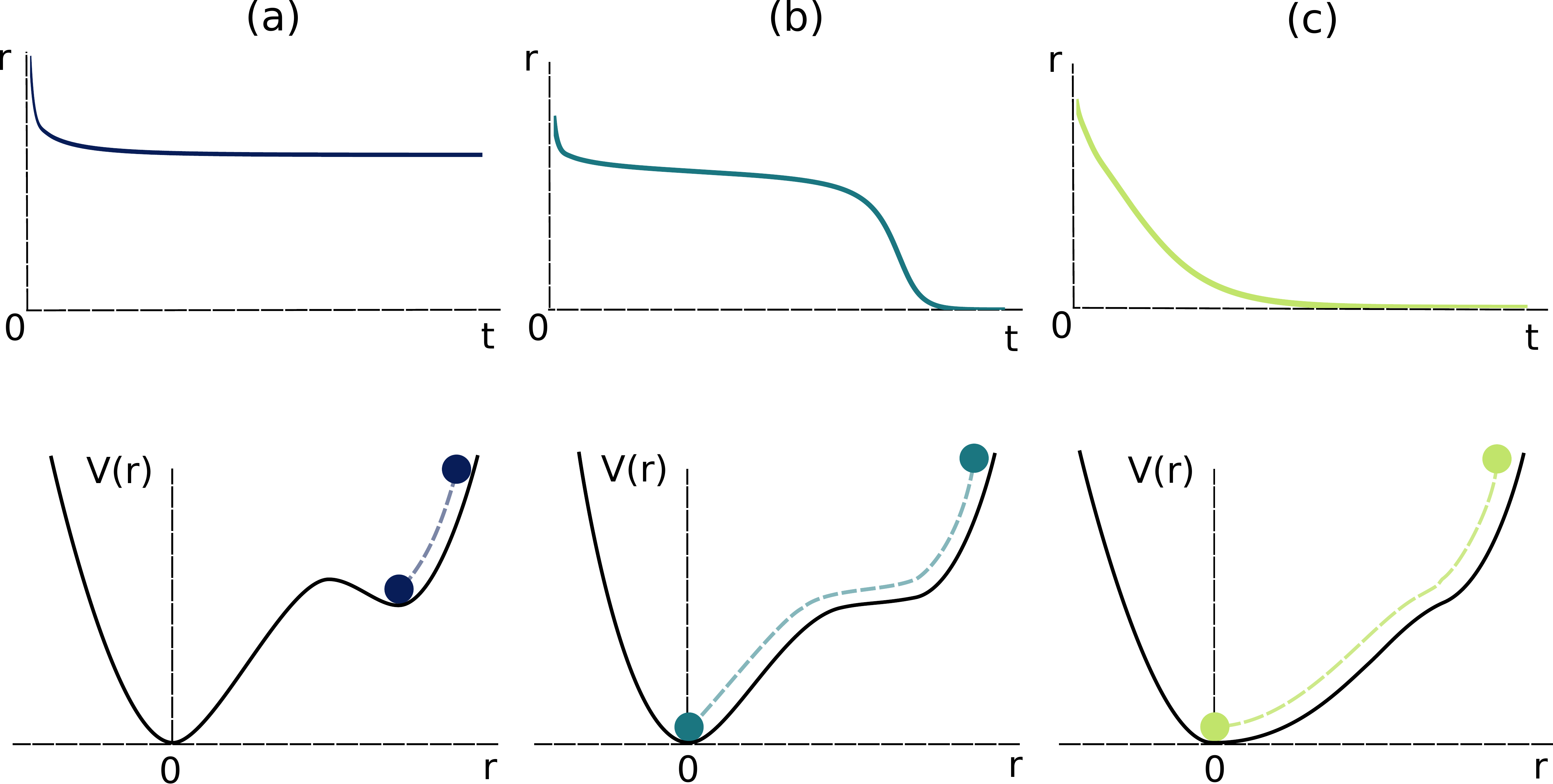}
		\vspace{5mm}
\caption{Effective potential model describing the evolution of $r=|\langle a \rangle|$ for (a) $\varepsilon>\varepsilon_c$ (b) $\varepsilon$ just below $\varepsilon_c$ (c) $\varepsilon$ well below $\varepsilon_c$. Each panel shows a cartoon of the time evolution of the order parameter in the system above a sketch of the corresponding effective potential which also indicates the initial and final values of $r$. }
\label{fig:mfpot}
\end{figure}
%

 Much of the behaviour of the order parameter seen in figure \ref{fig:vdpinit} can be captured using a simple classical effective-potential model\,\cite{Marcuzzi2014}, $V(r)$, such as that sketched in figure~\ref{fig:mfpot}. In such a model the dynamics of the order parameter is given by
 \begin{equation}
 \dot{r} = -\partial V/\partial r, \label{eq:vdot}
 \end{equation}
 and hence as soon as the gradient in the potential becomes zero $r$ will stop changing, giving rise to a fixed point in the dynamics.

Although we don't have a way of deriving the form of $V(r)$, its basic properties are clear. It will always have a stable fixed-point solution at $r=0$ (the unsynchronised state), and above a critical value of  $\varepsilon$ a second stable fixed-point solution with $r>0$ should emerge  [see figure~\ref{fig:mfpot}(a)], reflecting the coexistence of unsynchronised and synchronised ($r>0$) states. Furthermore, a third, unstable fixed point in the form of a peak between the two minima in the potential must emerge at the same time as the $r>0$ fixed point: the effective potential model undergoes a saddle-node bifurcation\,\cite{Strogatz1994} at a critical coupling. The unstable fixed point corresponds to the separatrix between synchronised and unsynchronised states seen in figure \ref{fig:vdpinit}. Figure~\ref{fig:mfpot}(b) shows the potential close to, but below, the critical value of $\varepsilon$. Here there is only one fixed point, at $r=0$, but the proximity to the critical point means that the gradient of the potential becomes very shallow, and correspondingly a bottleneck appears\,\cite{Strogatz1994} in the dynamics which means the time taken for $r$ to decay becomes extremely long. As figure \ref{fig:mfpot}(c) illustrates, for small enough $\varepsilon$ the long timescale decay is expected to disappear. One example of a potential which incorporates these features is
\begin{equation}
V(r) = \frac{1}{8\varepsilon}r^2- \frac{f(R)}{3}r^3 +\frac{g(R)}{4}r^4,
\label{eq:Vr}
\end{equation}
where the functions of $R$, $f$ and $g$, always take positive values.

This kind of effective potential model predicts the phase structure of the system by design, but it also predicts more subtle features of the system such as the scaling behaviour of the slow decay time that emerges for couplings below the critical value. We look in detail at the scaling behaviour of the relaxation time in the vdP system in section \ref{sec:criticalscaling} where we compare it to the prediction of the classical bifurcation model.

\section{Synchronised Versus Unsynchronised States}
\label{sec:flucs}
\begin{figure}
	\centering
		\includegraphics[width=1.0\textwidth]{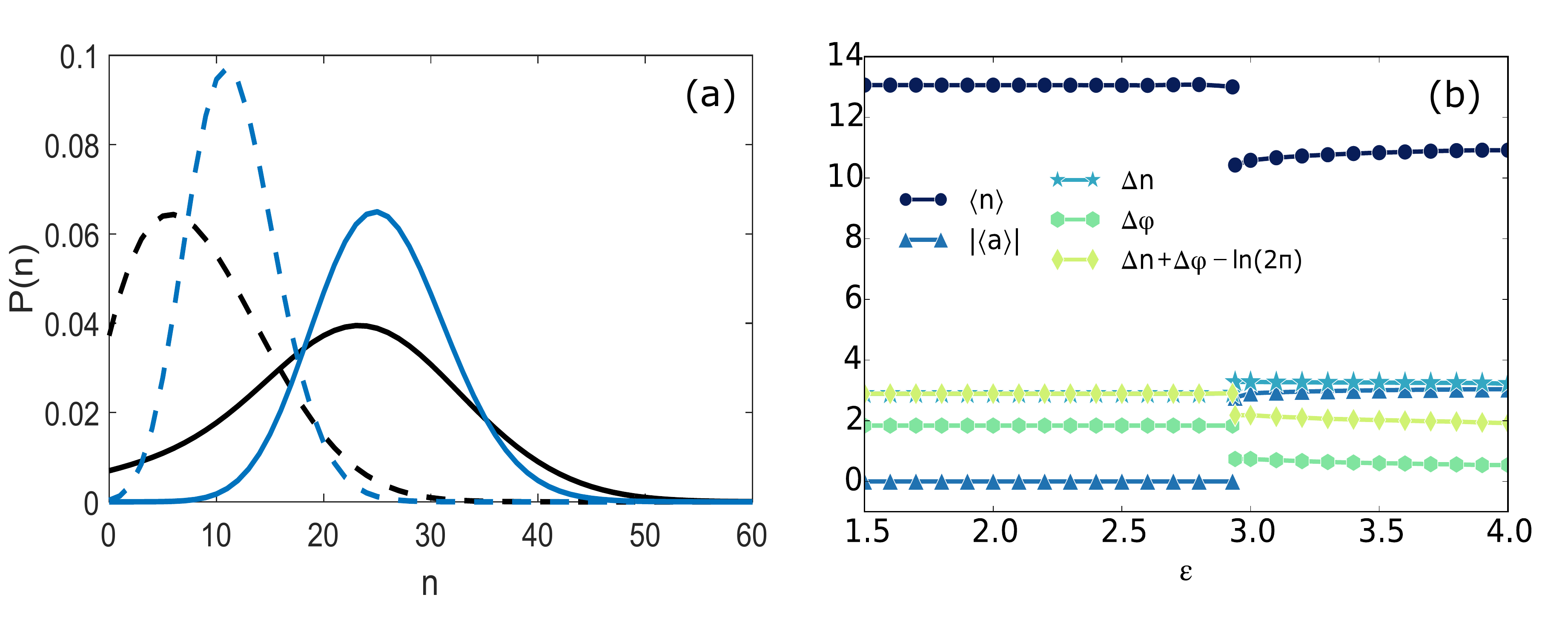}
\caption{Properties of synchronised and unsynchronised states. (a) Diagonal elements of the density operator, $P(n)=\langle n|\rho|n \rangle$, in the limit of long times for the unsynchronised state (blue curves) and examples of synchronised states (black curves); the dashed and full curves are for $R=0.02$ and $R=0.045$ respectively.  The synchronised states are obtained in the limit of long times for $\varepsilon=1.06$ (4.4) for $R=0.02$ (0.045),  just above the transition at  $\varepsilon_c\simeq1.057$ (4.331). (b) Long-time values of some key quantities in the system across the synchronisation transition. Here $R=0.04$  and the transition occurs at $\varepsilon=\varepsilon_c\simeq2.932$. For $\varepsilon<\varepsilon_c$ the system is always in the unsynchronised state (which is independent of $\varepsilon$); for $\varepsilon>\varepsilon_c$ the system reaches a synchronised state whose properties are weakly dependent on $\varepsilon$. }
\label{fig:one}
\end{figure}

We now turn to the properties of the synchronised and unsynchronised states which emerge in the limit of long times, beyond the value of the order parameter $r=|\langle a\rangle|$ that we have focussed on so far. The number distribution, $P(n)=\langle n|\rho|n\rangle$ (the diagonal elements of the density operator in the number state basis), shown in figure \ref{fig:one}(a) reveals clear differences between the synchronised and unsynchronised states which emerge in the limit of long times. The average occupation number, $\langle n\rangle$, is reduced in the synchronised state compared to the unsynchronised case, something which is not surprising as a reduction in the oscillation amplitude is a common accompaniment to synchronisation when classical limit-cycle oscillators are coupled together\,\cite{Pikovsky2003}. However, the $P(n)$ distribution is clearly much broader when the oscillators are synchronised and has a much larger vacuum state occupation probability. This suggests an interplay between phase and number fluctuations at the transition which is something we might expect intuitively. Figure \ref{fig:one}(a) also illustrates the way in which larger values of $R$ push the $P(n)$ distribution down to lower occupation numbers.

 To quantify how the number and phase properties of the quantum state of the vdP system change at the transition we need to adopt suitable measures of the uncertainty in both of these quantities. There are a number of different ways of defining phase uncertainty\,\cite{Iwo1993}, starting from the phase distribution
 \begin{equation}
 P(\varphi)=\frac{1}{2\pi}\langle \varphi|\rho|\varphi\rangle,
 \end{equation}
with
\begin{equation}
|\varphi\rangle=\sum_{n=0}^{\infty}{\textrm e}^{in\varphi}|n\rangle
\end{equation}
an eigenstate of the Susskind-Glogower operator $\sum_{n=0}^{\infty}|n\rangle\langle n+1|$\,\cite{knightbook}. Here we choose to work with an entropic measure of the phase uncertainty
\begin{equation}
\Delta \varphi = -\int_{-\pi}^\pi \textrm{d}\varphi P(\varphi) \textrm{ln} P(\varphi).
\end{equation}
This has the advantage that it and the corresponding number uncertainty,
\begin{equation}
\Delta n = -\sum_{n=0}^\infty P(n)\textrm{ln} P(n),
\end{equation}
together obey an uncertainty relation\,\cite{Iwo1993,coles}
\begin{equation}
\Delta\varphi + \Delta n \geq \textrm{ln}(2\pi).
\label{eq:varnvarphi}
\end{equation}
The lower bound of the uncertainty relation is reached for any pure number state (i.e.\ $\rho=|n=m\rangle\langle n=m|$ for $m=0,1,2\dots$) for which the phase distribution is flat $P(\varphi)=1/2\pi$ so $\Delta\varphi=\textrm{ln}(2\pi)$ and $\Delta n=0$.

Figure~\ref{fig:one}(b) compares the behaviour of the steady-state values of the uncertainties $\Delta n$ and $\Delta \varphi$ across the transition along with $|\langle a \rangle|$, $\langle n \rangle$ and $\Delta n + \Delta \varphi-\textrm{ln}(2\pi)$ [following equation~\ref{eq:varnvarphi}]. We know already that the synchronisation order parameter $|\langle a \rangle|$ is zero until the critical value of $\varepsilon$, at which point it takes a finite value (for an appropriate choice of initial conditions) indicating (partial) phase synchronisation. The phase uncertainty naturally drops at the transition point: it is maximal for the unsynchronised state $\Delta \phi=\textrm{ln}(2\pi)$ and must be lower when a phase preference emerges. The conjugate variable, the average occupation number $\langle n \rangle$, drops significantly at the transition and the uncertainty in $n$ goes up by a small amount. That a rise in number uncertainty should accompany a drop in phase uncertainty is perhaps not surprising, given their conjugate relationship, but the combined uncertainty in this case always remains well above the lower bound for the entropic uncertainties, $\textrm{ln}(2\pi)$. What is interesting here is that the phase-ordering that occurs at the transition is accompanied by number-disordering even when the latter is not required to ensure that the uncertainty relation is satisfied.

\section{Dynamics of Metastable Decay}
\label{sec:unlocking}

 In this section we look at the dynamics of the system just below the critical coupling where the system eventually evolves to a single unsynchronised state, but the time it takes to reach that state can become extremely large. We focus in particular on the question of how exactly the system makes its transition to the final unsychronised state. Then in the next section we will analyse the way in which the lifetime of the transient state scales with the distance from the critical coupling.

\begin{figure}
	\centering
		\includegraphics[width=1.0\textwidth]{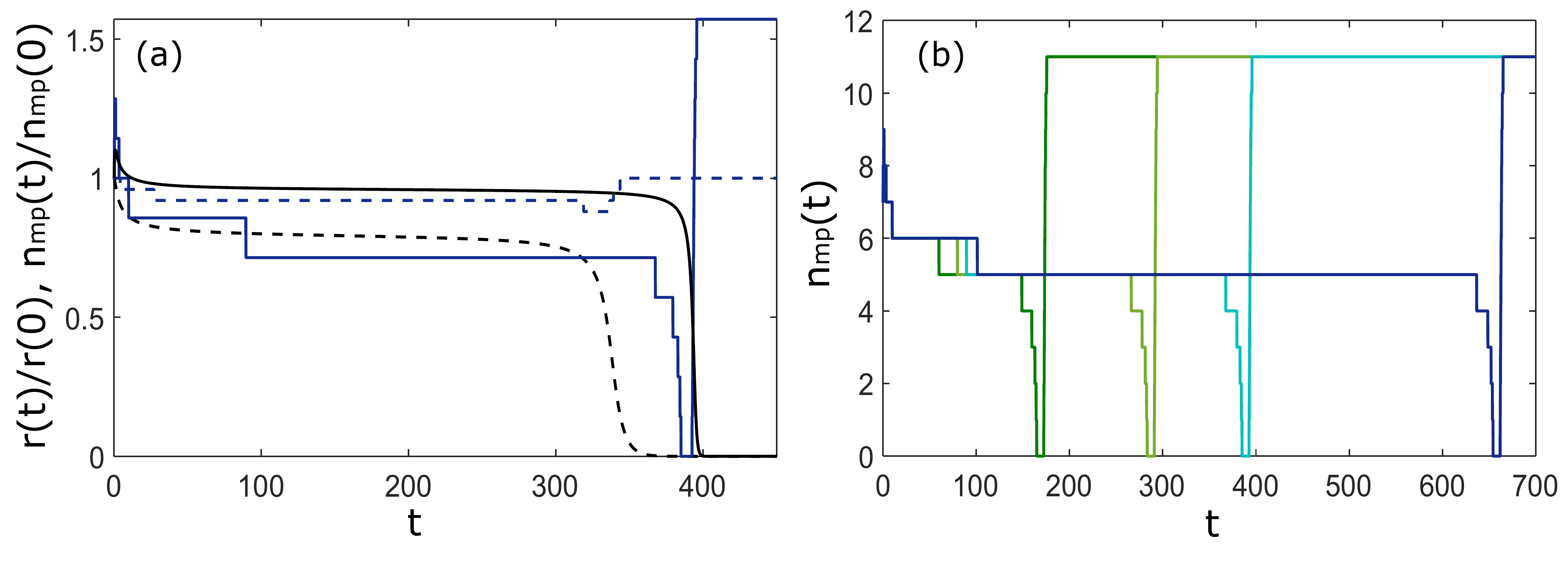}
\caption{Number and phase dynamics just below critical coupling. (a) Comparison of the dynamics of the order parameter $r=|\langle a\rangle|$ (black curves) and most probable number state $n_{mp}$ (blue curves) with $R=0.045$ and $R=0.02$ shown as full and dashed lines respectively (all quantities are normalised to their initial values). The coupling is $\varepsilon=1.055$ ($4.325$), just below the critical value $\varepsilon_c\simeq 1.057$ ($4.331$) for $R=0.02$ ($0.045$) (b) Evolution of $n_{mp}$ for a range of couplings ($\varepsilon=4.3,4.32,4.325,4.329$) just below $\varepsilon_c\simeq4.331$ for $R=0.045$. The initial state in each case is a coherent state with eigenvalue $\alpha=5$ $(2.6)$ for $R=0.02$ ($0.045$).}
\label{fig:dynamics}
\end{figure}

For couplings just below $\varepsilon_c$,  the order parameter exhibits a long period of very slow decay followed by an abrupt drop in its value. The final drop in $r$ is accompanied by rapid changes in the number distribution of the oscillator, which are clearly seen in the behaviour of $n_{mp}$, the $n$ value corresponding to the peak in the $P(n)$ distribution, as shown in figure \ref{fig:dynamics}(a). In contrast to the order parameter, the evolution of $n_{mp}$ is not monotonic. Instead it displays a kind of latching behaviour: as the value of $r$ drops abruptly $n_{mp}$ first dips before rising again to a value that is larger than its initial value.

Figure \ref{fig:dynamics}(a) also reveals an interesting dependence on the value of $R$. As expected from what we saw of the long-time steady states of the system [e.g.\, figure \ref{fig:one}], for smaller limit cycles (i.e.\ larger values of $R$) where the quantum noise is strong (and semiclassical descriptions provide a less accurate description of the system\,\cite{lee}), the slow relaxation process involves a substantial readjustment of both the number and phase properties of the system's state. Indeed, for smaller limit-cycles the value of $n_{mp}$ goes through very large variations---actually dropping to zero before growing again to a value larger than before. In contrast,  for larger limit-cycles (smaller $R$ values) there is only a modest dip in $n_{mp}$.

The drop and subsequent growth of $n_{mp}$ that occurs at the same time that $r$ undergoes rapid decay isn't sensitive to the overall length of the decay time. Figure \ref{fig:dynamics}(b) shows that the variation of $n_{mp}$ during the final rapid decay always takes the same form.

\begin{figure}
	\centering
		\includegraphics[width=0.75\textwidth]{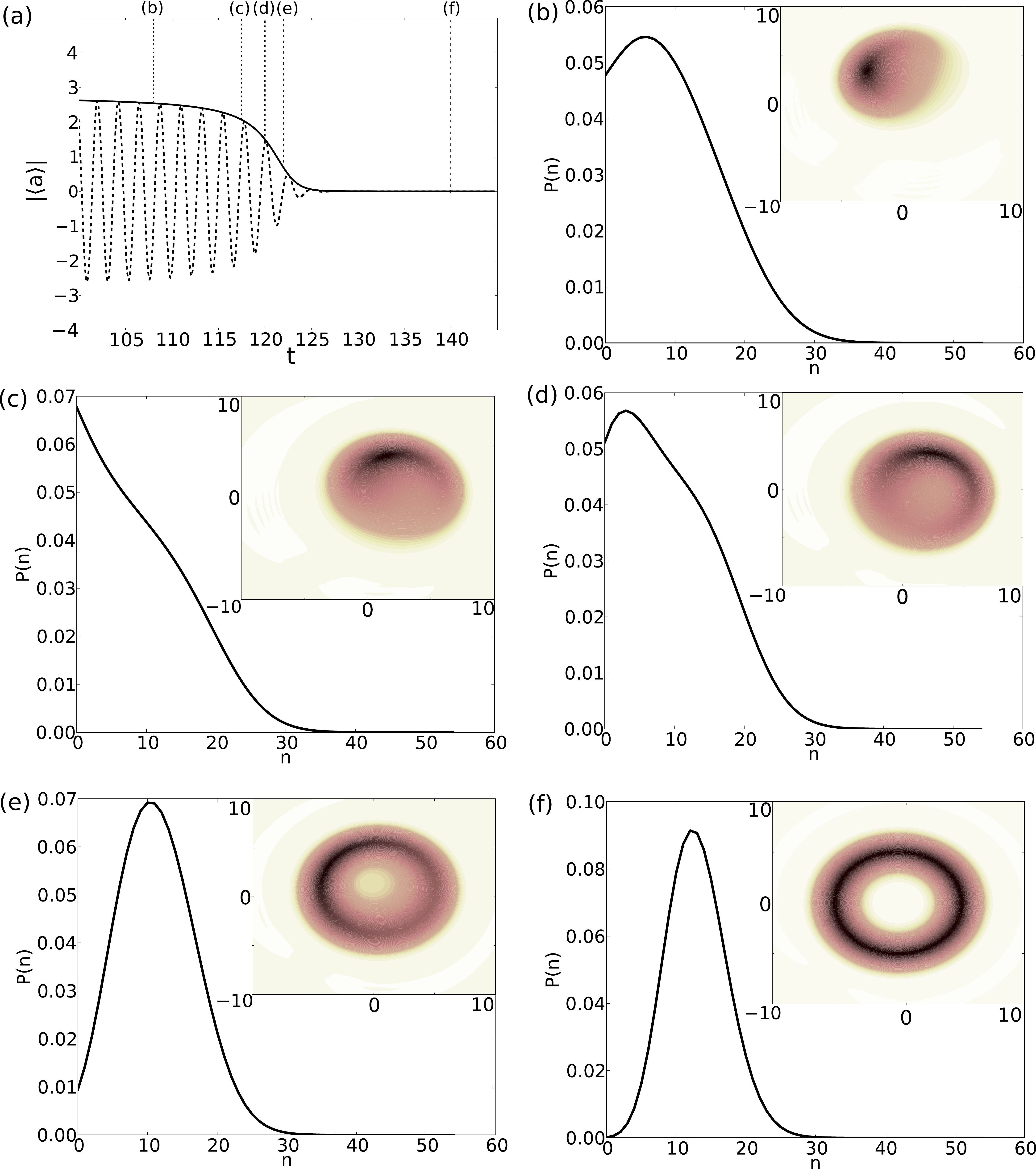}
\caption{Dynamics just below critical coupling. (a) Last stages of the decay in the order parameter $r=|\langle a\rangle|$ (full line), the oscillating real part of $\langle a\rangle$ is also shown (dashed line). (b)--(f) Photon number distribution, $P(n)$, and Wigner function (inset) for different times in the decay, indicated in (a). Here $R = 0.04$ and $\varepsilon=2.9$ ($\varepsilon_c\simeq2.932$). Note that the Wigner functions are positive throughout with darker colours indicating larger values.}
\label{fig:vdptrans}
\end{figure}

Figure~\ref{fig:vdptrans} provides a detailed illustration of the latching dynamics during the final rapid decay of the transient synchronised state. As the state decays, the number distribution $P(n)$ initially broadens and its peak drops to zero (for large enough $R$). The distribution then narrows as its peak moves to higher occupation numbers. The Wigner function\,\cite{knightbook} (which we calculated numerically using QuTiP\,\cite{qutip}) provides additional information about the phase during this transition. Throughout the slow decay stage it has a single well-defined peak centred away from the origin which indicates that the system has a preferred phase. During the final rapid decay, the Wigner function peak becomes smeared out around a central dip. Finally, the Wigner function reaches a limit-cycle state with no phase preference---matching the steady state of an equivalent uncoupled vdP oscillator\,\cite{lee}.

\section{Critical Scaling}
\label{sec:criticalscaling}

The critical scaling of the relaxation time in a classical dynamical system just below a saddle-node bifurcation (such as that described by the simple potential model given by \ref{eq:vdot} and \ref{eq:Vr})  takes the universal form\,\cite{Strogatz1994,Marcuzzi2014}
\begin{equation}
t\propto \left(1-\frac{\varepsilon}{{\varepsilon_c}}\right)^{b}, \label{eq:power}
\end{equation}
with $b=-0.5$. To see whether this matches the behaviour of our vdP system we looked in detail at the way in which the relaxation times of the order parameter grows with time as the coupling approaches the critical value.

\begin{figure}
	\centering
		\includegraphics[width=1.0\textwidth]{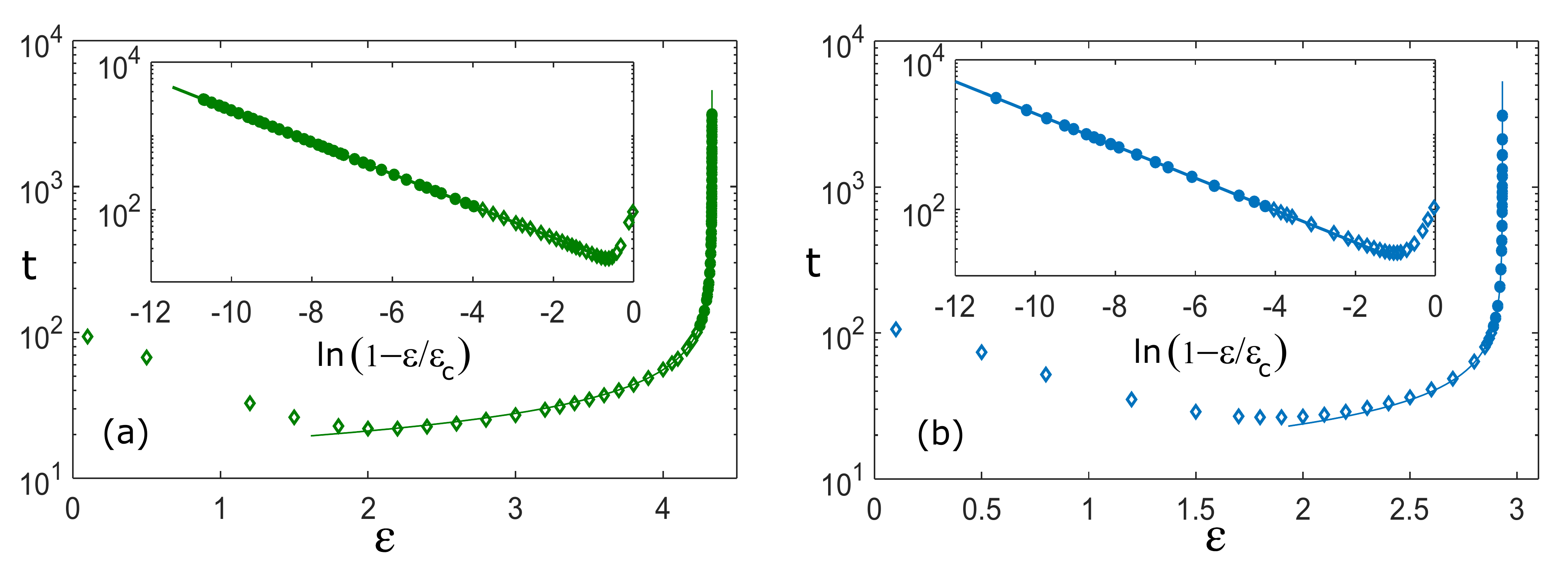}
\caption{Critical scaling of the relaxation time, $t$. The results of numerical integrations are shown as filled circles or empty diamonds for (a) $R=0.045$ (b) $R=0.04$. In each case, the line is a best-fit to the filled circles (see main text for details). For $R=0.045$ ($0.04$) we obtained $b=-0.4961$ $(-0.4946)$ with the best-fit provided by $\varepsilon_c=4.3311949$ ($2.9316769$). The numerical integrations used an initial coherent state with eigenvalue $\alpha=2.6$ ($2.75$) for $R=0.045$ ($0.04$) and  $t$ was defined as the time taken for $r$ to drop below $0.005$.}
\label{fig:cscale1}
\end{figure}

As figure \ref{fig:cscale1} shows, the critical scaling in the relaxation time of the vdP system for different values of $R$ is indeed consistent with the predictions of the classical bifurcation model. Since we had no {\emph{a priori}} knowledge of the precise value of the critical coupling strengths, we obtained critical exponents by {\emph{assuming}} the relation given by $\ref{eq:power}$. We then carried out linear fits for a range of choices of critical coupling, making no assumption about the value of the exponent $b$. The values of $\varepsilon_c$ used in figure \ref{fig:cscale1} are those for which the best fits to the assumed relation were obtained and the corresponding critical exponents were almost exactly $-0.5$ in each case. Since the relaxation behaviour for weak couplings is very different to that near the critical coupling (the relaxation time actually decreases with increasing coupling), we carried out our fits using only couplings for which the relaxation time was about at least as long as in the weak-coupling limit (the full circles in figure \ref{fig:cscale1}).

One very surprising feature of the dynamics is the wide range of couplings which seem to be well-described by \ref{eq:power}. Typically, one expects critical scaling to apply in a rather narrow range around the critical point, but figure \ref{fig:cscale1} shows that it can extend over a substantial range of coupling strengths. Furthermore, this range seems to increase with $R$, i.e.\ as the amplitude of the underlying limit-cycle gets smaller.

\section{Conclusion and Discussion}
\label{sec:conclusion}
We have explored the synchronisation dynamics of a model describing a large number of quantum vdP oscillators with all-to-all couplings. Adopting a quantum mean-field description, we carried out numerical integrations of the resulting effective non-linear quantum master equation. For couplings above a critical value the system evolves towards either a synchronised state or an unsynchronised state, depending on initial conditions. We found that these states differ not just in their phase properties, but also in the properties of their number distributions. In a sense the synchronisation transition can be thought of as involving both phase-ordering and number-disordering: the synchronised state has a well-defined phase, but its number distribution is always broader than that of the corresponding unsynchronised state which has a flat phase distribution.

We found that the dynamics of the system is rather rich with a number of interesting features. Just below the critical coupling the system always evolves towards the unsynchronised state, but the time taken to reach it can become extremely large. Looking at the dynamics of the slow relaxation process in detail we find that the system displays an interesting `latching' behaviour in which the average occupation number drops before rapidly rising again as the system approaches the unsynchronised state. The relaxation time for couplings below the critical value displays a scaling which is consistent with that predicted by a simple classical model describing a system undergoing a saddle-node bifurcation. However, in the regime where the occupation numbers of the system are relatively small (and the behaviour is not well-described by semiclassical models\,\cite{lee}), the scaling seems to apply over an unexpectedly broad range of couplings below the critical value.

Although we have focussed on a specific oscillator model involving vdP oscillators, we expect our results to apply rather generally to quantum limit-cycle oscillators coupled via a simple coherent coupling. For example, very similar results are found\,\cite{cdt} for many-body synchronisation in an analogous oscillator system which is instead based on the micromaser\,\cite{davis}.

 Our findings also suggest some promising directions for future work. It would be interesting to explore the dynamics within the vicinity of phase transitions in other non-equilibrium many-body quantum systems within the  mean-field approximation. It will also be worth investigating whether the dynamics is qualitatively similar in models in which the quantum mean-field approximation is relaxed (e.g. cluster mean-field descriptions based on plaquettes with two or more oscillators\,\cite{cmf}) or indeed in systems containing several rather than many oscillators (perhaps using models based on spin-systems which naturally have a much smaller state space\,\cite{roulet1,roulet2}). Finally, the question of why the behaviour of the many-body vdP system can end up being well described by critical scaling for a wide range of parameters is surely worth pursuing.

\section*{Acknowledgments}
 We thank J. P. Garrahan and M. Marcuzzi for very helpful discussions and are grateful for access to the University of Nottingham High Performance Computing Facility. CDT acknowledges funding in the form of a studentship from the EPSRC (EP/L50502X/1).

\end{document}